\documentclass[openacc]{rsproca_new}

\usepackage[numbers]{natbib}
\usepackage{enumitem}

\titlehead{Research}

\begin{document}

\title{Solar Atmospheric Abundances in Space \& Time}

\author{
S. L. Yardley$^{1,2,3}$
}

\address{$^{1}$Northumbria University, School of Engineering, Physics, and Mathematics, Ellison Place, NE1 8ST, UK \\
$^{2}$University College London, Mullard Space Science Laboratory, Holmbury St. Mary, Dorking, Surrey, RH5 6NT, UK \\
$^{3}$Donostia International Physics Center, Paseo Manuel de Lardizabal 4, 20018 San Sebasti{\'a}n, Spain \\

}

\subject{Astrophysics, observational astronomy, solar system, spectroscopy, elemental abundances}

\keywords{Elemental abundances, first ionisation potential, ponderomotive force, Alfv{\'e}n waves, solar corona, chromosphere, active regions, solar flares, magnetic activity, spectropolarimetry, solar--stellar connection}

\corres{S.~L.~Yardley\\
\email{steph.yardley@northumbria.ac.uk}}

\jname{rspa}
\Journal{Proc. R. Soc. A}

\begin{abstract}
Elemental abundances provide a powerful diagnostic of the physical mechanisms and processes that heat the solar atmosphere and drive the solar wind. The First Ionisation Potential (FIP) effect and its inverse (IFIP) are observed both on the Sun and other stars however, the underlying fractionation mechanisms, their dependence on the magnetic field topology, and the role of wave dynamics and turbulence in the chromosphere are not entirely understood. To address these challenges, a focused team, including observers, theorists, modellers and instrument scientists, spanning a range of career stages and institutions, came together for the Royal Society Theo Murphy meeting ``Solar Atmospheric Abundances in Space and Time". As a result of this meeting, the team worked in collaboration to produce 16 publications for this Special Issue. These publications are introduced here, including a discussion of the open questions and future directions in the context of advances in numerical modelling and current and upcoming solar and stellar missions.

This article is part of the Royal Society Theo Murphy Meeting Special Issue ``Solar Atmospheric Abundances in Space and Time''.

\end{abstract}
\maketitle



\section{Introduction}
Although all of the mass and energy that flows through the solar atmosphere and into the solar wind comes from the photosphere, these regions have different elemental abundances. This surprising fact is linked to the first ionisation potential (FIP) effect, where elements that are easily ionised in the chromosphere are enhanced in active regions and the solar wind, while difficult to ionise elements retain their photospheric abundances. The degree of fractionation is quantified by the FIP bias, which is defined as the ratio of coronal to photospheric elemental abundances, with values greater than one indicating the enhancement of low-FIP elements. The reverse (inverse FIP effect) is also observed on the Sun and more typically in very active M-dwarf stars, with both effects likely related to the coronal heating mechanism. 

One theory to explain the FIP and IFIP effects uses the separation of ions and neutrals by the ponderomotive force arising from the reflection and refraction of Alfv{\'e}n waves in the chromosphere \citep{Laming:2015}. Here, small-scale reconnection events in the corona generate downward propagating Alfv{\'e}n waves, giving rise to an upward ponderomotive force (FIP effect). In contrast, subchromospheric reconnection generates upward waves, leading to a downward ponderomotive force (IFIP effect). However, the details of this process remain poorly understood, and it is not clear that the hydrostatic nature of the model can accurately capture realistic solar dynamics or the role of transient phenomena (which pass through the critical region too quickly to produce a FIP effect). Further recent advances include state-of-the-art multi-fluid 2.5-D radiative MHD simulations \citep{martinez-sykora:2023}. Using this approach, large ponderomotive forces are associated with regions of enhanced magnetic field and highly energetic phenomena. In contrast to the Laming model, a multi-fluid approach produces Alfv{\'e}n wave damping and ponderomotive forces in the direction of wave propagation (i.e., Alfv{\'e}n waves propagating upward induce an upward ponderomotive force). Additionally, \citep{Reville:2021} uses a shell-turbulence model to investigate FIP fractionation along both open and closed magnetic field lines, coupling Alfv{\'e}n wave-driven turbulence to an analytical fractionation model that also holistically deals with coronal heating and solar wind acceleration. This study showed that turbulence alone can create a ponderomotive force strong enough to produce a low-FIP bias in the chromosphere and transition region, without the resonant Alfv{\'e}n waves central to the Laming model. 

Although EUV spectroscopic measurements of coronal elemental abundances showing the FIP effect have been available from the Hinode mission since 2006 \citep{Brooks:2011,Brooks:2015}, attempts have only recently been made to relate these to wave activity \citep{Baker:2021} and theoretical model predictions \citep{Mihailescu:2023}. In addition, the inverse FIP effect has only recently been discovered on the Sun \citep{Doschek:2015} and in the solar wind \citep{Brooks:2022}. 

While these significant recent advances have greatly improved our understanding of this fundamental physical process, they have also resulted in a myriad of open questions. As an acknowledgement of this, it was decided to apply for funding to support a dedicated multi-day meeting where the leading experts in observations, radiative transfer processes, and simulations could gather alongside the leading proponents of the best current models. The aims of this meeting were to enable different theoretical predictions to be confronted with observations in a collegiate and open discussion, to consolidate recent progress in this topic, and ultimately to try and provide a roadmap for future research avenues. 

\section{Meeting Overview}

Funding was provided by The Royal Society Hooke Committee via the award of a Theo Murphy meeting (\url{https://royalsociety.org/science-events-and-lectures/scientific/scientific-meetings/}), enabling a two day meeting which took place at The Apex Grassmarket Hotel in Edinburgh from 16-17 June 2025. A total of 31 participants including 15 invited speakers from around the world attended the meeting for an intensive two days of discussion and debate. As a result of this meeting, a series of publications including those focused on recent scientific results, perspectives on recent progress, and future opportunities were solicited for a Special Issue of Philosophical Transactions of the Royal Society A. 

When identifying the invited speakers for the Theo Murphy meeting, a large emphasis was placed on ensuring a diverse programme, covering a range of career stages, geographical locations, and gender balance. Due to restrictions on space, the meeting was strictly invite-only, with applicants requesting an invitation via the meeting website. Of the 31 participants who ultimately attended the meeting, 35\% were female, with a total of 11 countries represented and a career spectrum ranging from undergraduate students to Professors.

\begin{figure}
    \centering
    \includegraphics[width=1\linewidth]{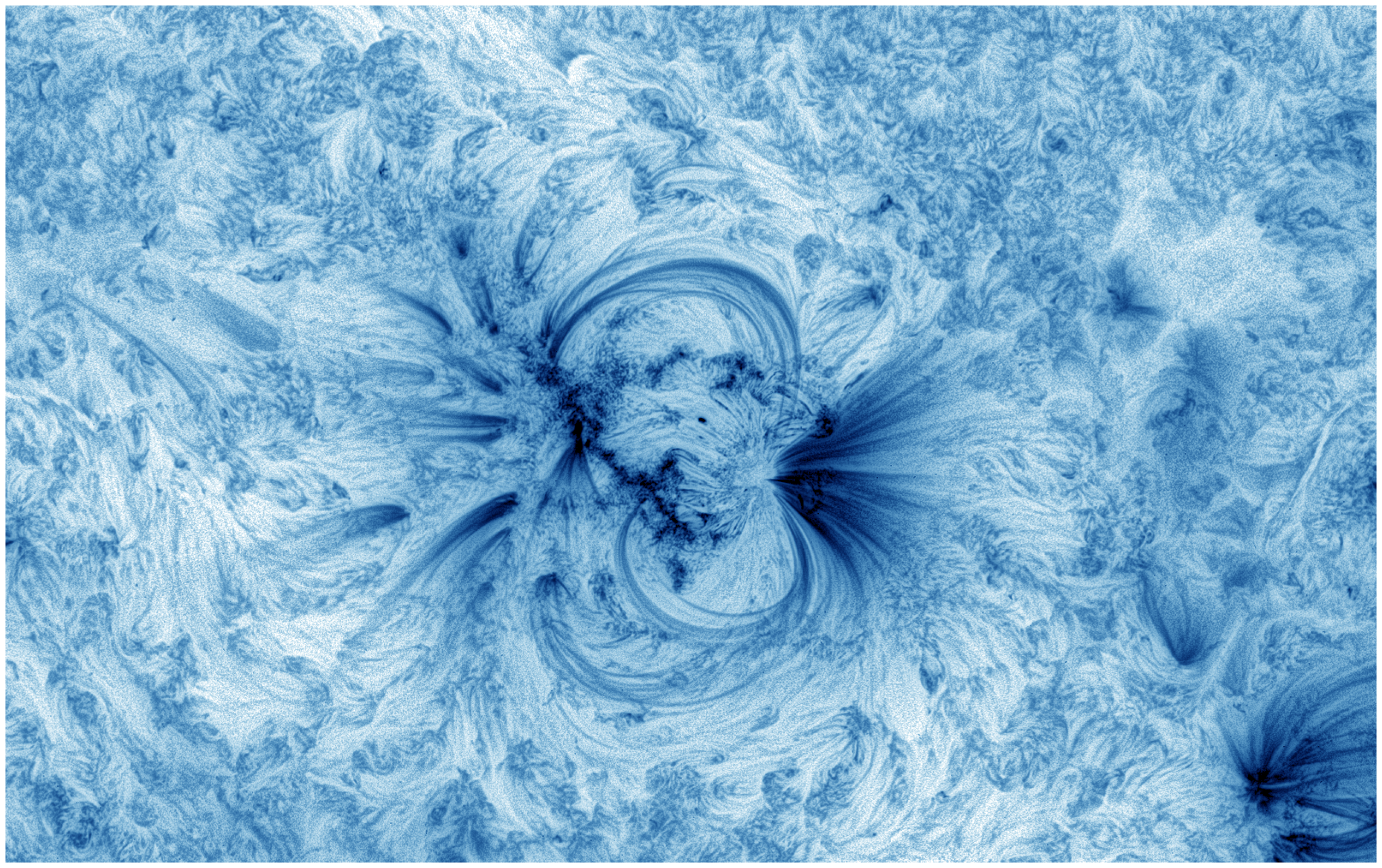}
    \caption{The front cover image of the Special Issue. A 171~$\AA\ $ image of active region 12546 taken by the Atmospheric Imaging Assembly onboard the Solar Dynamics Observatory on 20/05/2016, that has been enhanced by applying a Multi-scale Gaussian technique \citep{Morgan:2014}.}
    \label{fig:cover}
\end{figure}

\section{Publications}

The Royal Society Theo Murphy Meeting ``Solar Atmospheric Abundances in Space and Time'' resulted in a Special Issue of the Philosophical Transactions of the Royal Society A, consisting of 16 publications (not including this introduction). Notably, almost a third of these manuscripts are led by undergraduate and PhD students, reflecting the meeting's inclusion of early career researchers. The Special Issue was first open for submissions in May 2024, closing in January 2026, and overseen by the Guest Editors Dr David Long, Dr David Brooks and Dr Deborah Baker.

\subsection{Pondermotive Force Models}
The Special Issue begins with a reflection on the ponderomotive force model by Laming et al., which illustrates how the magnetic field geometry of the solar corona influences the reflection and refraction of chromospheric Alfv{\'e}n waves, leading to different scenarios of FIP fractionation, and demonstrating the effect of wave resonance through a coronal loop model. Resonant waves reproduce the abundance patterns seen in the remote-sensing observations of the corona, which includes the behaviour of Sulphur as a high-FIP element while, non-resonant low-frequency waves produce Sulphur fractionation more consistent with in situ measurements of the solar wind. These results reinforce the idea that compositional variations offer a diagnostic of wave excitation mechanisms and coronal dynamics.

While Laming et al.\ focus on the FIP effect, Martinez-Sykora et al.\
investigate the ponderomotive force in the case of the inverse FIP (IFIP) effect, going beyond semi-empirical models and using a 1D multi-fluid MHD model. The authors show that a negative ponderomotive acceleration can arise when strong magnetic fields and flux tube expansion with height counteract multi-fluid dissipation and damping. This is consistent with observations of IFIP at the locations of light bridges present in strong, complex sunspots, offering an alternative explanation to subsurface reconnection and may also help interpret IFIP signatures seen in highly active and late spectral type stars.

Building on the question of how fractionated plasma evolves once in the corona, Reep et al.\ use hydrodynamic simulations to investigate how elemental abundances vary in solar flares, assuming a low-FIP enhancement at loop footpoints and triggering impulsive heating events. The authors find that the coronal abundance response depends on the depth at which fractionation occurs. Sharply peaked enhancements produce a localised anomaly near the loop apex that can trigger coronal rain, while broader enhancements yield a uniformly fractionated corona without coronal rain. These results predict an important correlation between coronal rain formation, the strength of flare heating and fractionation depth, which could be tested using observations from current and upcoming missions such as Hinode/EIS and IRIS, and MUSE and Solar-C/EUVST, respectively.

Brooks et al.\ continue this theme by dynamically modelling the variation of elemental abundances in radiative hydrodynamic simulations. The authors start from an IFIP-dominated corona, as observed in active M-dwarf stars, then evaporate photospheric plasma into post-flare loops, finding that the resulting solution resembles the solar case of a low-FIP enhanced corona. The authors also show that the depletion of low-FIP elements reduces cooling at the loop apex, suppressing coronal condensation and rain formation during M-dwarf flares. Such behaviour may be detectable in EUV or X-ray observations with abundance variations already being identified in XRISM and NICER spectra, while TESS light curves may reveal coronal rain signatures in the flare late phase.

\subsection{Elemental Abundance Observations}

The first attempts of whether coronal composition anomalies can be traced back to specific chromospheric properties are explored by Testa et al.\ The authors analysis reveals that there is a possible increase of microturbulence in the chromosphere and non-thermal line widths in the transition region corresponding with areas of strong enhancement of low-FIP elements but the picture is not concrete, particularly in the vicinity of sunspots. These results pose several broader questions surrounding the physical mechanism responsible across different magnetic field environments and on what spatial and temporal scales these mechanisms operate. A deeper understanding requires both advancements in fractionation models and deriving tighter constraints from observations of different atmospheric layers. This motivates the upcoming Solar-C/EUVST mission that would overcome current observation limitations due to its wavelength coverage and high throughput along with the ability to simultaneously observe the atmosphere from the chromosphere to the corona. 

Taking a complementary approach to linking chromospheric wave behaviour with coronal composition, Murabito \& Stangalini use full Stokes measurements of the Ca II 854.2~nm line from IBIS at the Dunn Solar Telescope along with Hinode/EIS observations of the corona to investigate the diagnostic potential of chromospheric spectropolarimetry in order to probe wave behaviour associated with FIP fractionation. The authors focus on active region 12546 (see Figure~\ref{fig:cover}), which is an exceptionally large and long-lived active region, whose stability allows wave dynamics to be isolated from its evolution. Enhanced oscillations in the 4-6 mHz band are found on one side of the sunspot umbra, which coincides with regions of strong coronal FIP bias that has been previously linked to magnetic oscillations, consistent with wave reflection in the ponderomotive force model. However, it is evident that there is a lack of coordinated ground-based chromospheric spectropolarimetric and space-based coronal observations, which makes it challenging to understand the variability of the elemental abundances between the corona and the photosphere. 

Along similar lines, Mesoraca et al.\ analyse active region 12665 over a full disk passage with 16 Hinode/EIS composition maps and SDO/HMI Dopplergrams. The authors find recurring 4-5 mHz power enhancements in the sunspot umbra that appear to correlate to changes in the coronal FIP bias, although current observational constraints limit depth of the analysis. This again motivates the upcoming Solar-C/EUVST and MUSE missions that, along with DKIST, which will provide the spectroscopic sensitivity and temporal coverage required to quantify this correlation.

Although previous papers focus on the FIP effect, Baker et al.\ review the current understanding of the IFIP effect, presenting a detailed analysis of active region 11967 by combining Hinode/EIS, IRIS, SDO/AIA, and Fermi observations with IRIS 2+ inversions. The authors find that highly localised IFIP plasma can be explained by torsional Alfv{\'e}n waves generated below the chromospheric fractionation region, which pinpoints the importance of processes occurring below the chromosphere. Significant progress is hindered by the rarity of solar IFIP observations, the absence of direct chromospheric diagnostics for IFIP plasma, and the lack of simultaneous multi-layer observations. These are gaps where Solar-C/EUVST, DKIST, and advances in multi-fluid modelling are needed. Expanding on both the observational detections and the modelling framework is essential not only for understanding localised patches of IFIP on the Sun but also for interpreting the IFIP-dominated coronae of active M-dwarf stars.

Focusing on the same active region studied by Baker et al., Orlovskij et al.\ examine the fractionation and behaviour of Sulphur, an intermediate-FIP element, which is at the low/high-FIP boundary, using observations from Hinode/EIS combined with parameters derived from potential field source surface (PFSS) model. By using different diagnostic pairs from the Hinode/EIS observations the authors show that the low-FIP elements, including Sulphur, decrease above a mean field strength of $\sim$150$~G$ with little dependence on the coronal loop length. Multi-fluid MHD simulations offer a qualitative explanation through downward ponderomotive forces acting in strong-field regions such as the active region core, but whether this threshold is physical or reflects model limitations requires a larger observational sample.

Next, Spruksta et al.\ compare three Hinode/EIS FIP bias diagnostics across an active region and regions of quiet Sun, finding distinct variations between diagnostics that reflect their differing sensitivities to plasma properties and derivation methods (e.g. DEM vs intensity ratios). The authors argue that quoting fixed FIP bias values for a region is misleading and instead recommend that distribution quartiles should be reported. Going forward, this is a more complete approach that Solar-C/EUVST and ongoing work with Solar Orbiter will help refine.

From FIP bias diagnostics to the physical evolution, the perspective by Brooks et al.\ draws on the connections between the timescales of coronal abundance evolution across active region lifetimes, the solar cycle and stellar evolution to discuss similarities in these domains. The authors propose a picture where the dissipation of magnetic complexity controls the competition between heating processes during active region evolution and potentially explains the observed variations in coronal composition. The perspective extends the picture to solar cycle and stellar evolutionary timescales. The authors also highlight some methodological challenges inherent in differential emission measure-based abundance analysis, differences in species fractionation (including the behaviour of Sulphur also discussed in this issue by Orlovskij et al.), and the reliance on assumed solar photospheric abundances for stellar measurements, thus identifying key areas where further investigation is needed. 

Motivated by the question of how plasma properties vary with active region evolution, Power et al.\ utilise IRIS full-disk spectroheliograms to compare chromospheric and transition region plasma signatures across active regions at different evolutionary stages. Although the C II, Si IV, and Mg II lines show no clear variability between the regions, distinct differences emerge in the Mg II k/h ratio. The regions with the highest FIP bias exhibit double-peaked profiles suggestive of variable plasma density, which could influence the local wave propagation. These results provide useful constraints on previously underexplored spectral diagnostics, with future work combining higher-cadence observations and modelling needed to clarify the relationship between plasma properties and fractionation.

While the previous papers rely on established diagnostic techniques, Zambrana-Prado et al.\ develop a telemetry-efficient diagnostic for measuring the relative abundances of Sulphur to Nitrogen using Solar Orbiter/SPICE. The authors apply an optimised Linear Combination Ratio technique, which requires only four EUV lines, compared to the several lines needed for DEM inversions. The method enables routine composition mapping using observations that are already regularly acquired by the Solar Orbiter/SPICE instrument, with the potential to deliver higher-level data products that would be accessible to non-specialists. This diagnostic is particularly promising for probing the transition region and the polar regions as Solar Orbiter reaches higher latitudes, and for linking remote-sensing plasma composition to in situ solar wind measurements. 

Mzerguat et al. utilise Solar Orbiter/SPICE observations and the Sulphur to Nitrogen diagnostic of Zambrana-Prado et al.\ to investigate the fractionation present in coronal hole plumes. Both plumes the authors investigated show Sulphur fractionation located within strong magnetic footpoints, in contrast to the surrounding interplume plasma, providing evidence consistent with chromospheric wave dynamics as a driver within the ponderomotive force model. In order to resolve the fine-scale structures requires coordinated high-resolution observations that span the chromosphere to corona, for example, combining future missions such as Solar-C/EUVST with current DKIST observations and EUI high-resolution imaging from Solar Orbiter.

Returning to modelling, To et al.\ combine hydrodynamic simulations from HYDRAD with ponderomotive force calculations using FIPpy, a new open-source code, in order to explore how chromospheric dynamics influence elemental fractionation to move beyond the static atmospheres assumed in existing works. The authors show that when acoustic wave flux drops below a critical threshold, mass-dependent thermal velocities dominate over the ponderomotive force, which produces unexpected fractionation patterns, while chromospheric turbulence acts to suppress the FIP bias. These results potentially explain the abundance variations observed during solar flares. Extending FIPpy to stellar parameter space offers a route towards predictive modelling of the FIP/IFIP as a function of stellar type and activity.

\section{Conclusion and Future Directions}

The contributions to this Special Issue highlight both the significant progress and also persistent challenges in understanding solar and stellar elemental abundances and their evolution in space and time. There are a number of recurring themes that emerge, including the role of the ponderomotive force model in explaining both the FIP and IFIP fractionation patterns, the importance of intermediate-FIP elements such as Sulphur being utilised as sensitive diagnostics of the underlying fractionation processes, and the growing recognition that elemental abundances contain information about chromospheric dynamics, wave propagation, magnetic topology, and their interaction across spatial and temporal scales. Across these scales, connections are drawn between the patterns in solar and stellar abundances from active region evolution through to the solar cycle and to the IFIP-dominated coronae of M-dwarf stars, showcasing the broad astrophysical reach of the work in this Special Issue.

A consistent conclusion reached in many of the manuscripts included in this Special Issue is that current observational capabilities remain a limiting factor. Multi-layer observations spanning from the photosphere through the chromosphere to the corona are essential but rare, and the lack of simultaneous high-resolution spectropolarimetric and Doppler diagnostics restricts the ability to distinguish between wave modes and constrain fractionation models. Coordinated observing campaigns between existing facilities, for example, combining Hinode/EIS coronal composition maps with IRIS chromospheric and transition region diagnostics, DKIST spectropolarimetry and SDO/AIA EUV images would already address many of these gaps, with several publications demonstrating the scientific return even if coordination across some but not all facilities is achieved. Furthermore, Solar Orbiter/SPICE probes layers that are distinct from those accessible to Hinode/EIS and is opening avenues for complementary composition analysis through intermediate-FIP elements such as Sulphur. The telemetry-efficient techniques presented in this Special Issue will become particularly valuable as Solar Orbiter reaches higher latitudes during the declining phase of the solar cycle, when the polar coronal holes form. 

X-ray missions including XRISM have begun to reveal abundance variability in stellar flares, while TESS optical light curves contain signatures of coronal rain in the flare late phase. Looking ahead, Solar-C/EUVST will deliver the broad wavelength coverage and spectroscopic sensitivity needed to probe composition simultaneously from the chromosphere to the corona, while MUSE will provide spatially resolved coronal spectroscopy at unprecedented cadence. Together with strengthened coordination of current facilities, these upcoming capabilities will directly address the observational gaps identified throughout this Special Issue.

Advances in modelling and theory are equally as important as the observations. Several contributions provide a good foundation, but static atmospheric assumptions are insufficient. Single-fluid or equilibrium models cannot capture chromospheric dynamics, multi-fluid effects, and the balance between ponderomotive acceleration and turbulence, which all interact to shape the resulting abundance patterns. Open-source tools that have been developed such as FIPpy represent a step forward in self-consistent treatments that couple fractionation with atmospheric evolution, while hydrodynamic simulations of flare-driven evaporation reveal the sensitivity of coronal composition to the depth and spatial extent of the fractionation occurring at loop footpoints. Extending these modelling efforts into the stellar parameter space offers a promising route in terms of a unified physical picture of the FIP/IFIP effect. 

The publications in this Special Issue collectively demonstrate that understanding solar atmospheric abundances requires a concerted, interdisciplinary effort that combines remote-sensing and in situ observations across wavelengths, advanced numerical modelling and theory, and attention to the methodology of plasma diagnostics. The Royal Society Theo Murphy meeting was a necessary first step as it brought together observers, modellers, and instrument scientists across all career stages and serves as a foundation as the next generation of space-borne missions, ground-based telescopes, and models come online. We look forward to the advances this community will make towards a unified understanding of elemental fractionation across solar and stellar atmospheres.

\section{Key Open Questions}
The publications in this Special Issue highlight a number of key open questions that are central to advancing our understanding on the topic of solar and stellar abundances. These questions range from a better understanding of the underlying fractionation mechanisms, the diagnostics used to determine elemental abundances, and the connections between the Sun, the solar wind and other stars. While these questions have been grouped into three broad themes, many of the questions are interconnected and progress in one theme will likely lead to progress in the others.\\

{\bf Fractionation Mechanisms}
\begin{enumerate}[label=(\arabic*)]
\item Does FIP fractionation arise primarily from Alfv{\'e}n waves generated in the corona or turbulence driven by photospheric motions, and can chromospheric observations distinguish between these generation mechanisms?
\item Can the dominating mechanism between chromospheric turbulence, acoustic wave flux, and ponderomotive acceleration determine the abundance variations observed across different coronal structures, timescales, and solar phenomena?
\item Why are IFIP observations so rare on the Sun, and what determines their localisation within complex active regions?
\item Is subchromospheric reconnection or multi-fluid wave damping the dominant IFIP driver, or do both operate and does it depend upon the magnetic field environment?\\
\end{enumerate}

{\bf Elemental Abundance Diagnostics}
\begin{enumerate}[label=(\arabic*), resume]
\item What determines the behaviour of Sulphur along open and closed magnetic fields, and how does this relate to the underlying wave dynamics?
\item How sensitive are DEM-based abundance measurements to assumed photospheric abundances, and what are the implications for comparing FIP bias across solar and stellar observations?
\item On what timescales does coronal composition respond to changes in heating and magnetic complexity, and how does the depth and spatial extent of fractionation influence the resulting coronal dynamics?\\
\end{enumerate}

{\bf Solar Wind and Solar–Stellar Connection}
\begin{enumerate}[label=(\arabic*), resume]
\item To what extent does the magnetic field topology control the degree of FIP fractionation, and what does this imply for solar wind formation models?
\item What will high-latitude composition mapping of polar coronal holes reveal about fast wind fractionation and solar wind formation?
\item What determines whether a star exhibits FIP or IFIP as a function of stellar type and activity level, and can fractionation models developed for the Sun be extended to the stellar context?
\end{enumerate}

The discussions and publications resulting from this Theo Murphy meeting represent the first step in addressing these questions, and we look forward to continued collaboration with the broader community to make further progress.

\appendix

\section{List of Acronyms}

\begin{table}[ht]
\centering
\caption{A list of the mission or instrument acronyms used in this manuscript.}
\label{tab:acronyms}
\begin{tabular}{@{}p{2cm}p{8cm}@{}}
\hline
\textbf{Acronym} & \textbf{Definition} \\
\hline
SDO & Solar Dynamics Observatory \\
IRIS & Interface Region Imaging Spectrograph \\
DKIST & Daniel K. Inouye Solar Telescope \\
XRISM & X-Ray Imaging and Spectroscopy Mission \\
TESS & Transiting Exoplanet Survey Satellite \\
EIS & EUV Imaging Spectrometer (Hinode) \\
HMI & Helioseismic and Magnetic Imager (SDO) \\
AIA & Atmospheric Imaging Assembly (SDO) \\
EUI & Extreme Ultraviolet Imager (Solar Orbiter) \\
SPICE & Spectral Imaging of the Coronal Environment (Solar Orbiter) \\
IBIS & Interferometric BIdimensional Spectropolarimeter (DST) \\
EUVST & EUV High-Throughput Spectroscopic Telescope (Solar-C) \\
MUSE & Multi-slit Solar Explorer \\
\hline
\end{tabular}
\end{table}

\enlargethispage{20pt}

\dataccess{This article has no additional data.}
\aucontribute{S.L.Y. attended the RS Theo Murphy meeting, acted as a reviewer, and wrote this introduction for the Special Issue of the Philosophical Transations of the Royal Society A: "Solar Atmospheric Abundances in Space and Time".}
\competing{The author declares that they have no competing interests.}
\funding{This work was supported by the Science Technology and Facilities Council award ST/X003787/1.}
\ack{S.L.Y. would like to acknowledge the Science Technology and Facilities Council for the award of an Ernest Rutherford Fellowship (ST/X003787/1). The Guest Editors are grateful to the Royal Society and its staff, particularly Harry Cutmore, Ellen Porter, Ashleigh Carver, and Amy Simons for their guidance, advice and patience when helping plan, organise, and run the Theo Murphy meeting and editing the current Special Issue.}


\vskip2pc

\bibliographystyle{RS} 

\bibliography{bibliography} 

\end{document}